\documentclass[conference,10pt]{IEEEtran}
\IEEEoverridecommandlockouts

\usepackage{cite}
\usepackage{amsmath,amssymb,amsfonts}
\usepackage{graphicx}
\usepackage{textcomp}
\usepackage{xcolor}
\def\BibTeX{{\rm B\kern-.05em{\sc i\kern-.025em b}\kern-.08em
		T\kern-.1667em\lower.7ex\hbox{E}\kern-.125emX}}

\usepackage{subfigure}
\usepackage{epstopdf}

\begin{document}

\title{Over-the-Air Computation with Imperfect Channel State Information}

\author{\IEEEauthorblockN{Yilong Chen\textsuperscript{*}, Guangxu Zhu\textsuperscript{\dag}, and Jie Xu\textsuperscript{*}}
	\IEEEauthorblockA{\textsuperscript{*}SSE and FNii,
	The Chinese University of Hong Kong (Shenzhen), Shenzhen, China \\
	\textsuperscript{\dag}Shenzhen Research Institute of Big Data, Shenzhen, China \\
	Email: 118010033@link.cuhk.edu.cn, gxzhu@sribd.cn, xujie@cuhk.edu.cn}
}

\maketitle

\begin{abstract}
This paper investigates the effect of imperfect channel state information (CSI) on the over-the-air computation (AirComp) system, in which multiple wireless devices (WDs) send individual messages to one access point (AP) for distributed functional computation. By particularly considering the channel estimation errors, we jointly design the transmit coefficients at the WDs and the receive strategy at the AP, for minimizing the computation mean squared error (MSE). First, we consider the single-input single-output (SISO) case with each WD and AP equipped with one single antenna, in which the globally optimal solution to the computation MSE minimization problem is obtained in closed form. Next, we consider the single-input multiple-output (SIMO) case with multiple receive antennas at the AP, in which a high-quality solution is obtained based on alternating optimization and convex optimization. For both cases, the optimized power control solution at the WDs follows a threshold-based regularized channel inversion structure; while for the SIMO case, the receive beamforming at the AP follows a sum-minimum MSE (MMSE) structure. It is shown that with finite receive antennas, a non-zero computation MSE is inevitable due to the channel estimation errors even when the WDs' transmit powers become infinity; while with massive receive antennas, a vanishing MSE is achievable when the channel vectors are independent and identically distributed. Finally, numerical results are provided to demonstrate the effectiveness of the proposed designs. 
\end{abstract}


\section{Introduction}

Over-the-air computation (AirComp) has been recognized as a new multiple access technique towards beyond fifth-generation (B5G) and six-generation (6G) wireless networks to facilitate distributed data aggregation for various applications such as federated edge learning and distributed sensing \cite{zhu2021over}. Different from conventional multiple access techniques focusing on data delivery, AirComp aims to compute function values of distributed data from separate wireless devices (WDs). By exploiting the superposition property of multiple access channel (MAC) together with proper preprocessing at the WD transmitters and postprocessing at the access point (AP) receiver, AirComp is able to compute a class of so-called nomographic functions such as arithmetic mean, weighted sum, geometric mean, polynomial, and Euclidean norm \cite{goldenbaum2014nomographic}.

In general, AirComp can be realized in both coded and uncoded manners \cite{zhu2021over}. Recently, the uncoded AirComp has attracted growing research interests (see, e.g., \cite{gastpar2008uncoded, cao2020optimized, zhang2021gradient, cao2022optimized, cao2022transmission, zhu2018mimo, zhai2021hybrid, yu2020optimizing, jung2021performance, zhu2021one}) due to its optimality in Gaussian MAC with independent and identically distributed (IID) sources (in terms of minimizing the computation mean squared error (MSE)) \cite{gastpar2008uncoded} and its simplicity in implementation. For instance, the authors in \cite{cao2020optimized} studied the transmit power control design for minimizing the average computation MSE in fading channels, by properly balancing the tradeoff between the signal misalignment and noise-induced errors. Such design was then extended to the over-the-air federated edge learning systems to accelerate the convergence of training machine learning models \cite{zhang2021gradient, cao2022optimized, cao2022transmission}. Furthermore, by considering the multi-antenna setup, the authors in \cite{zhu2018mimo} presented the framework of multiple-input multiple-output (MIMO) AirComp for multimodal data aggregation, and those in \cite{zhai2021hybrid} studied the design of hybrid analog and digital beamforming for massive MIMO AirComp. Notice that the practical implementation of power control and beamforming in AirComp highly depends on the availability of channel state information (CSI) at transceivers, and those prior works normally assumed perfect CSI to facilitate the transceiver design.

In practice, the CSI can be obtained at the AP and the WDs via channel estimation by exploiting the channel reciprocity, which, however, can induce channel estimation errors \cite{yoo2006capacity} that may degrade the AirComp performance. In the literature, there have been a handful of prior works analyzing the effect of imperfect CSI on the AirComp performance under different setups with, e.g., intelligent reflecting surface (IRS)-assisted cloud radio access network (C-RAN) \cite{yu2020optimizing}, unmanned aerial vehicles (UAVs) \cite{jung2021performance}, and federated edge learning \cite{zhu2021one}. Nevertheless, how to optimize the AirComp transceiver design by taking into account the imperfect CSI has not been well investigated yet, thus motivating our work in this paper. 

This paper investigates the uncoded AirComp system, in which multiple WDs simultaneously transmit uncoded data to one single AP for distributed functional computation. We study the optimized transceiver design in the presence of channel estimation errors, with the objective of minimizing the computation MSE, subject to the maximum power constraints at individual WDs. First, in the single-input single-output (SISO) case with one single antenna at each WD and AP, we obtain the closed-form globally optimal solution to the computation MSE minimization problem. Then, in the single-input multiple-output (SIMO) case with multiple antennas at the AP, we obtain a high-quality solution by proposing an efficient algorithm based on alternating optimization and convex optimization. In both cases, the optimized power control policy at WDs follows a threshold-based regularized channel inversion structure, where the regularization depends on the channel estimation errors. In the SIMO case, the optimized receive beamforming at the AP follows a sum-minimum MSE (MMSE) structure. In addition, it is shown that with finite receive antennas at the AP, a non-zero computation MSE is inevitable due to the channel estimation errors even when the transmit powers at WDs go to infinity; while with massive receive antennas, a vanishing MSE is achievable when the channel vectors are IID, thus showing the benefit of using massive antennas to mitigate the channel estimation errors in AirComp. Finally, numerical results are provided to demonstrate the impact of channel estimation errors on the computation MSE, and also validate the effectiveness of our proposed designs as compared to existing benchmarks.

\textit{Notations}: Bold lower-case letters are used for vectors. For
a vector \(\boldsymbol{a}\), \(\boldsymbol{a}^*\), \(\boldsymbol{a}^H\), and \(\|\boldsymbol{a}\|\) denote its conjugate, conjugate transpose, and Euclidean norm, respectively. \(\boldsymbol{I}\) denotes the identity matrix whose dimension will be clear from the context. \(\mathbb{C}^{m\times n}\) denotes the \(m \times n\) dimensional complex space. \(\mathbb{E}[\cdot]\) denotes the statistic expectation.

\section{System Model and Problem Formulation}

In this paper, we consider an uncoded AirComp system, in which an AP aims to compute the function value of distributed data from a set \(\mathcal{K} \triangleq \{1,\dots, K\}\) of \(K \ge 1\) WDs. It is assumed that the AP is equipped with \(N_r \ge 1\) receive antennas and each WD is equipped with one single transmit antenna. Let \(s_k\) denote the transmit message by WD \(k \in \mathcal{K}\), where \(s_k\)'s are independent random variables with zero mean and unit variance.
The AP is interested in computing the averaging function of \(s_k\)'s,\footnote{Our proposed designs are extendible to other nomographic functions via proper pre-processing and post-processing \cite{goldenbaum2014nomographic}.} i.e., \(
f = \frac{1}{K} \sum_{k=1}^K s_k.
\)

Let \(\boldsymbol{h}_k \in \mathbb{C}^{N_r \times 1}\) denote the channel vector from WD \(k \in \mathcal{K}\) to the AP, \(b_k\) denote the transmit coefficient at WD \(k\). The received signal at the AP is given by
\begin{equation}
\boldsymbol{y} = \sum_{k=1}^K \boldsymbol{h}_k b_k s_k + \boldsymbol{z},
\end{equation}
where \(\boldsymbol{z} \in \mathbb{C}^{N_r \times 1}\) denotes the additive white Gaussian noise (AWGN) at the AP that is a circularly symmetric complex Gaussian (CSCG) random vector  with zero mean and covariance \(\sigma_z^2 \boldsymbol{I}\). Let \(P_k\) denote the maximum transmit power budget at WD \(k \in \mathcal{K}\). Accordingly, we have \(\mathbb{E}[|b_k s_k|^2] = |b_k|^2 \le P_k, \forall k \in \mathcal K\).

We consider that the AP only accesses imperfect CSI to coordinate the transceiver design, due to the channel estimation errors. Let \(\boldsymbol{\widehat{h}}_k \in \mathbb{C}^{N_r \times 1}\) denote the estimated channel vector for WD \(k\). Then we have \cite{yoo2006capacity} 
\begin{equation}
\boldsymbol{\widehat{h}}_k = \boldsymbol{h}_k + \boldsymbol{e}_k, \forall k \in \mathcal{K},
\end{equation}
where \(\boldsymbol{e}_k \in \mathbb{C}^{N_r \times 1}\) denotes the channel estimation error that is a CSCG random vector with zero mean and covariance \(\sigma_{e,k}^2 \boldsymbol{I}\).

Based on the estimated CSI \(\{\boldsymbol{\widehat{h}}_k\}\), the AP adopts the receive beamforming vector \(\boldsymbol{w} \in \mathbb{C}^{N_r \times 1}\) for data aggregation. Accordingly, the received signal is expressed as
\begin{equation}
\boldsymbol{w}^H \boldsymbol{y} = \sum_{k=1}^K \boldsymbol{w}^H (\boldsymbol{\widehat{h}}_k - \boldsymbol{e}_k) b_k s_k + \boldsymbol{w}^H \boldsymbol{z},
\end{equation}
which is used to recover the average function as \(
\widehat{f} = \frac{\boldsymbol{w}^H \boldsymbol{y}}{K}.
\)

We use the computation MSE as the performance metric of AirComp, which is expressed as follows to characterize the distortion of \(\widehat{f}\) with respect to the groundtruth average \(f\). 
\begin{equation}
\begin{aligned}
&\mathrm{MSE} = \mathbb{E} [(\widehat{f}-f)^2] \\
&= \frac{1}{K^2} \sum_{k=1}^K (|\boldsymbol{w}^H \boldsymbol{\widehat{h}}_k b_k - 1|^2 + \|\boldsymbol{w}\|^2 \sigma_{e,k}^2 |b_k|^2) + \|\boldsymbol{w}\|^2 \sigma_z^2),
\end{aligned} \label{MSE}
\end{equation}
where the expectation is taken over the randomness of both \(\{s_k\}\) and \(\{\boldsymbol{e}_k\}\).

Our objective is to minimize \(\mathrm{MSE}\) in \eqref{MSE} by jointly optimizing the transmit coefficients \(\{b_k\}\) at the WDs and the receive beamforming vector \(\boldsymbol{w}\) at the AP, subject to the individual power budgets at the WDs. The computation MSE minimization problem with channel estimation errors is formulated as problem (P1) in the following, where the constant coefficient \(\frac{1}{K^2}\) in \eqref{MSE} is dropped for notational convenience. 
\begin{equation*}
\begin{aligned}
(\text{P}1): \min_{\{b_k\}, \boldsymbol{w}} &\
\underbrace{\sum_{k=1}^K |\boldsymbol{w}^H \boldsymbol{\widehat{h}}_k b_k - 1|^2}_{\textrm{Signal misalignment error}}  + \underbrace{\sum_{k=1}^K \|\boldsymbol{w}\|^2 \sigma_{e,k}^2 |b_k|^2}_{\textrm{CSI-related error}} \\
&+ \underbrace{\|\boldsymbol{w}\|^2 \sigma_z^2}_{\textrm{Noise-induced error}} \\
\mathrm{s.t.} &\ |b_k|^2 \le P_k, \forall k \in \mathcal{K}.
\end{aligned}
\end{equation*}
The objective function in (P1) consists of three terms, including the signal misalignment error, the noise-induced error, and the CSI-related error (due to channel estimation errors). This is different from prior studies with perfect CSI (e.g., \cite{cao2020optimized}), where only the first two terms are taken into account. Also, due to the coupling between \(\{b_k\}\) and \(\boldsymbol{w}\) in the objective function, (P1) is non-convex, which is thus difficult to be optimally solved. Sections III and IV will deal with (P1) for the SISO and SIMO cases with \(N_r = 1\) and \(N_r > 1\), respectively.

\section{Optimal Transceiver Design for SISO Case with \(N_r=1\)}

This section considers the computation MSE minimization problem (P1) with \(N_r = 1\), which is reexpressed as 
\begin{equation*}
	\begin{aligned}
		(\text{P}2): \min_{\{b_k\}, w \ge 0} &\
		\sum_{k=1}^K (|w \widehat{h}_k b_k - 1|^2 + w^2 \sigma_{e,k}^2 |b_k|^2) + w^2 \sigma_z^2 \\
		\mathrm{s.t.} &\ |b_k|^2 \le P_k, \forall k \in \mathcal{K},
	\end{aligned}
\end{equation*}
where the combining vector \(\boldsymbol{w}\) in (P1) becomes a nonnegative-valued denoising factor \(w\) without loss of optimality. For problem (P2), with phase alignment, the optimality is achieved by setting \(b_k = \widetilde{b}_k \frac{\widehat{h}_k^*}{|\widehat{h}_k|}\), where \(\widetilde{b}_k \ge 0\) denotes the transmit amplitude of WD \(k \in \mathcal{K}\). Thus, (P2) is further simplified as
\begin{equation*}
	\begin{aligned}
		(\text{P}2.1): \min_{\{\widetilde{b}_k\}, w \ge 0} &\
		\sum_{k=1}^K \big((w |\widehat{h}_k| \widetilde{b}_k - 1)^2 + w^2 \sigma_{e,k}^2 \widetilde{b}_k^2\big) + w^2 \sigma_z^2 \\
		\mathrm{s.t.} &\ 0 \le \widetilde{b}_k \le \sqrt{P_k}, \forall k \in \mathcal{K}.
	\end{aligned}
\end{equation*}

Now, we deal with problem (P2.1). First, we optimize \(\{\widetilde{b}_k\}\) under given \(w \ge 0\). In this case, the optimization of \(\{\widetilde b_k\}\) can be decomposed into the following \(K\) subproblems (by dropping the constant term \(w^2 \sigma_z^2\)), each for one WD \(k \in \mathcal{K}\):
\begin{equation*}
	\begin{aligned}
		(\text{P}3. k):\min_{0 \le \widetilde{b}_k \le \sqrt{P_k}} &\
		(w |\widehat{h}_k| \widetilde{b}_k - 1)^2 + w^2 \sigma_{e,k}^2 \widetilde{b}_k^2. \\
	\end{aligned}
\end{equation*}
By checking the first derivative of the objective function, the optimal solution to problem (P3.\(k\)) is obtained as
\begin{equation}
	\widetilde{b}_k^* = \min \big(\sqrt{P_k}, \frac{|\widehat{h}_k|}{w (|\widehat{h}_k|^2 + \sigma_{e,k}^2)}\big), \forall k \in \mathcal{K}. \label{p_k^star}
\end{equation}

Next, we optimize \(w\). By substituting \eqref{p_k^star} in (P2.1), the optimization of \(w\) becomes 
\begin{equation*}
	\begin{aligned}
		(\text{P}4): \min_{w \ge 0} & \sum_{k=1}^K \Big(\big(\min(w \sqrt{P_k} |\widehat{h}_k| - 1, -\frac{\sigma_{e,k}^2}{|\widehat{h}_k|^2 + \sigma_{e,k}^2})\big)^2 \\
		&+ \big(\min(w \sqrt{P_k} \sigma_{e,k}, \frac{|\widehat{h}_k| \sigma_{e,k}}{|\widehat{h}_k|^2 + \sigma_{e,k}^2})\big)^2\Big) + w^2 \sigma_z^2.
	\end{aligned}
\end{equation*}
To solve problem (P4), we adopt the divide-and-conquer approach, by dividing \(\{w \ge 0\}\) into \(K+1\) intervals as
\begin{equation}
	\begin{aligned}
		\mathcal{F}_k &\triangleq \{w | 1/\rho_k > w \ge 1/\rho_{k+1}\}, k \in \{0\} \cup \mathcal{K}, \label{F}
	\end{aligned}
\end{equation}
where we define \(\rho_k \triangleq \sqrt{P_k} \frac{|\widehat{h}_k|^2 + \sigma_{e,k}^2}{|\widehat{h}_k|}\) as the channel quality indicator for WD \(k \in \mathcal{K}\), \(\rho_0 \to 0\), and \(\rho_{K+1} \to \infty\).

Then, we optimize \(w\) over each interval \(\mathcal{F}_k, k\in \{0\} \cup \mathcal K\), and compare their correspondingly achieved objective values to find the optimal solution. Notice that the optimization of \(w\) over \(\mathcal{F}_k, k \in \{0\} \cup \mathcal{K}\), in (P4) is given by
\begin{equation*}
	\begin{aligned}
		(\text{P}4. k): \min_{w \in \mathcal{F}_k} &\ F_k(w) \triangleq \sum_{i=1}^k \big((w \sqrt{P_i} |\widehat{h}_i| - 1)^2 + w^2 P_i \sigma_{e,i}^2\big) \\
		&+ \sum_{j=k+1}^K \frac{\sigma_{e,j}^2}{|\widehat{h}_j|^2 + \sigma_{e,j}^2} + w^2 \sigma_z^2,
	\end{aligned}
\end{equation*}
for which the optimal solution is
\begin{equation}
	w_k^* = \max \big(1/\rho_{k+1}, \min (\widetilde{w}_k, 1/\rho_k)\big),
\end{equation}
where \(\widetilde{w}_k = \frac{\sum_{i=1}^k \sqrt{P_i} |\widehat{h}_i|}{\sum_{i=1}^k P_i (|\widehat{h}_i|^2 + \sigma_{e,i}^2) + \sigma_z^2}\) corresponds to the solution to the equation \(\frac{d F_k(w)}{d w} = 0\).

By comparing the achieved values \(F_k(w_k^*)\)'s of (P4.\(k\))'s, the optimal solution of \(w\) to (P2) is given as
\begin{equation}
	w^* = w^*_{k^*}, k^* = \arg \min_{k \in \{0\} \cup \mathcal{K}} F_k(w_k^*). \label{w_SISO}
\end{equation}
Based on \eqref{w_SISO} together with \eqref{p_k^star}, the optimal solution of \(\{b_k\}\) to (P2) is finally obtained as
\begin{equation}
	b_k^* =
	\begin{cases}
		\sqrt{P_k} \frac{\widehat{h}_k^*}{|\widehat{h}_k|}, k \in \{1,\dots,k^*\}, \\
		\frac{\widehat{h}_k^*}{w^* (|\widehat{h}_k|^2 + \sigma_{e,k}^2)}, k \in \{k^*+1,\dots,K\}.
	\end{cases} \label{b_SISO}
\end{equation}

It is observed that the optimal transmit coefficient solution (or power control policy) in \eqref{b_SISO} follows a threshold-based regularized channel inversion structure. For each WD \(k \in \{1,\dots,k^*\}\) with poor channel quality and/or limited transmit power (i.e., \(\rho_k < 1/w^*\)), the full power transmission is applied; while for each WD \(k \in \{k^*+1,\dots,K\}\) with good channel quality and/or sufficient transmit power (i.e., \(\rho_k \ge 1/w^*\)), the regularized channel inversion power control is applied, where the regularization depends on the channel estimation error \(\sigma_{e,k}^2\). 

It is also interesting to analyze the computation MSE when each WD has asymptotically high transmit power (i.e., \( P_k \to \infty, \forall k\in\mathcal K\)), for which we have the following proposition.

\textit{Proposition 1}: When \(P_k \to \infty, \forall k \in \mathcal{K}\), it follows that \(\mathrm{MSE} \to \frac{1}{K^2} \sum_{k=1}^K \frac{\sigma_{e,k}^2}{|\widehat{h}_k|^2 + \sigma_{e,k}^2}\). 

\textit{Proof}: In this case, it follows from \eqref{w_SISO} and \eqref{b_SISO} that \(k^* = 0\), \(w^* \to 0\), and \(b_k^* = \frac{\widehat{h}_k^*}{w^* (|\widehat{h}_k|^2 + \sigma_{e,k}^2)}, \forall k \in \mathcal{K}\). By substituting them in \eqref{MSE}, we have \(\mathrm{MSE} \to \frac{1}{K^2} \sum_{k=1}^K \frac{\sigma_{e,k}^2}{|\widehat{h}_k|^2 + \sigma_{e,k}^2}\).

Proposition 1 shows that due to the existence of channel estimation errors \(\{\sigma_{e,k}^2\}\), a non-zero computation MSE becomes inevitable even when the transmit powers at WDs become infinity. This is different from the case with perfect CSI (i.e., \(\{\sigma_{e,k}^2 = 0\}\)), in which \(\mathrm{MSE} \to 0\) when \(P_k \to \infty, \forall k \in \mathcal{K}\). 

\section{Proposed Transceiver Design for SIMO Case with \(N_r > 1\)}

This section considers problem (P1) in the general case with \(N_r > 1\). To deal with the coupling of the transmit coefficients \(\{b_k\}\) and the receive combining vector \(\boldsymbol{w}\) in this case, we propose an efficient solution based on alternating optimization, where \(\{b_k\}\) and \(\boldsymbol{w}\) are updated alternately with the other given.

First, we optimize \(\{b_k\}\) in (P1) under given \(\boldsymbol{w}\). This corresponds to solving the following \(K\) subproblems, each for one WD \(k \in \mathcal{K}\):
\begin{equation*}
(\text{P}5. k): \min_{0 \le |b_k| \le \sqrt{P_k}}\
|\boldsymbol{w}^H \boldsymbol{\widehat{h}}_k b_k - 1|^2 + \|\boldsymbol{w}\|^2 \sigma_{e,k}^2 |b_k|^2.
\end{equation*}
Similarly as in problem (P2) and by replacing \(w |\widehat{h}_k|\) in \eqref{p_k^star} as \(\boldsymbol{w}^H \boldsymbol{\widehat{h}}_k\), the optimal solution to problem (P5.\(k\)) is obtained as \(b_k^\star = \widetilde{b}_k^\star \frac{\boldsymbol{\widehat{h}}_k^H \boldsymbol{w}}{|\boldsymbol{w}^H \boldsymbol{\widehat{h}}_k|}\) with
\begin{equation}
\widetilde{b}_k^\star = \min(\sqrt{P_k}, \frac{|\boldsymbol{w}^H \boldsymbol{\widehat{h}}_k|}{|\boldsymbol{w}^H \boldsymbol{\widehat{h}}_k|^2 + \|\boldsymbol{w}\|^2 \sigma_{e,k}^2}), \forall k \in \mathcal{K} \label{b^star}.
\end{equation}

The optimized transmit coefficient or equivalently power control solution in \eqref{b^star} is observed to follow a similar threshold-based regularized channel inversion power control as in \eqref{b_SISO}. For the WDs with sufficient transmit power and/or good channel quality (by viewing \(|\boldsymbol{w}^H \boldsymbol{\widehat{h}}_k|^2\) as the equivalent channel power gain), their transmit powers follow a regularized channel inversion structure; for the other WDs with limited transmit power and/or poor channel quality, the full power transmission is adopted. 

Next, we optimize \(\boldsymbol{w}\) in (P1) under given \(\{b_k\}\). This corresponds to solving the following unconstrained convex optimization problem:
\begin{equation*}
\begin{aligned}
(\text{P}6): \min_{\boldsymbol{w}} &\sum_{k=1}^K \big(|\boldsymbol{w}^H \boldsymbol{\widehat{h}}_k b_k - 1|^2 + \|\boldsymbol{w}\|^2 \sigma_{e,k}^2 |b_k|^2\big) \\
&+ \|\boldsymbol{w}\|^2 \sigma_z^2.
\end{aligned}
\end{equation*}
By setting the gradient of the objective function to be zero, we have the optimal solution to problem (P6) as
\begin{equation}
\boldsymbol{w}^\star = \big({\sum_{k=1}^K |b_k|^2 (\boldsymbol{\widehat{h}}_k \boldsymbol{\widehat{h}}_k^H + \sigma_{e,k}^2 \boldsymbol{I}) + \sigma_z^2 \boldsymbol{I}}\big)^{-1} \sum_{k=1}^K \boldsymbol{\widehat{h}}_k b_k \label{w^star}.
\end{equation}
The optimized receive beamforming solution in \eqref{w^star} is observed to have a sum-MMSE structure. This is in order to better aggregate the signals from all the WDs to facilitate the functional computation.

In summary, the alternating-optimization-based algorithm for solving (P1) is implemented in an iterative manner. In each iteration, we first obtain the transmit coefficients as \(\{b_k^\star\}\) in \eqref{b^star} (by solving (P5.\(k\))'s) under given \(\boldsymbol{w}\), and then update the receive beamforming vector as \(\boldsymbol{w}^\star\) based on \eqref{w^star} (by solving (P6)) under given \(\{b_k\}\). Notice that in each iteration problems (P5.\(k\))'s and (P6) are both optimally solved, and as a result, the updated computation MSE is ensured to be monotonically nonincreasing. As the computation MSE in (P1) is lower bounded, the convergence of our proposed alternating-optimization-based algorithm can be guaranteed. 

%

It is interesting to discuss the computation MSE in the cases with sufficient transmit powers or a massive number of receive antennas, for which we have the following two propositions.

\textit{Proposition 2}: Under any given receive beamforming vector \(\boldsymbol{w}\), if \(P_k \to \infty, \forall k\in\mathcal K\), then we have \(\mathrm{MSE} \to \frac{1}{K^2} \sum_{k=1}^K \frac{\|\boldsymbol{w}\|^2 \sigma_{e,k}^2}{|\boldsymbol{w}^H \boldsymbol{\widehat{h}}_k|^2 + \|\boldsymbol{w}\|^2 \sigma_{e,k}^2} \ge \frac{1}{K^2} \sum_{k=1}^K \frac{\sigma_{e,k}^2}{\|\boldsymbol{\widehat{h}}_k\|^2 + \sigma_{e,k}^2}\).

Proposition 2 can be similarly verified as Proposition 1, where the last inequality holds based on the Cauchy-Schwarz inequality. It follows from Proposition 2 that with a finite number of receive antennas, a non-zero MSE becomes inevitable even the WDs employ extremely high transmit powers.

\textit{Proposition 3}: 
If \(N_r \to \infty\) and \(\boldsymbol{h}_i\)'s (and equivalently \(\boldsymbol{\widehat{h}}_i\)'s) are IID random vectors, then we have \(\mathrm{MSE} \to 0\). 

\textit{Proof}: In this case, the channel vectors among different WDs become asymptotically orthogonal, i.e., \(\frac{1}{N_r} \boldsymbol{\widehat{h}}_i \boldsymbol{\widehat{h}}_j^H \approx \boldsymbol{0}\) and \(\boldsymbol{\widehat{h}}_i \boldsymbol{\widehat{h}}_i^H \approx N_r \sigma_h^2 \boldsymbol{I}\), \(\forall i,j \in \mathcal{K}, i \neq j\), where \(\sigma_h^2\) denotes the variance \(\boldsymbol{\widehat{h}}_i\)'s. Accordingly, we have \(\boldsymbol{w}^* = \frac{1}{\alpha+\beta+\gamma} \sum_{k=1}^K \boldsymbol{\widehat{h}}_k b_k\), where \(\alpha = \sum_{k=1}^K |b_k|^2 N_r \sigma_h^2\), \(\beta = \sum_{k=1}^K |b_k|^2 \sigma_{e,k}^2\), and \(\gamma = \sigma_z^2\). Substituting \(\boldsymbol{w}^*\) into \eqref{MSE}, the computation MSE becomes \(\frac{(\beta+\gamma)^2 + \alpha\beta + \alpha\gamma}{K^2 (\alpha+\beta+\gamma)^2}\). Hence, as \(N_r \to \infty\), we have  \(\alpha \to \infty\), and \(\mathrm{MSE} \to 0\).

Proposition 3 shows that increasing the number of receive antennas is efficient to combat against the imperfect CSI, thus showing the benefit of massive antennas in AirComp. 

\section{Numerical Results}

This section evaluates the AirComp performance of our proposed designs, in terms of the computation MSE. We consider the following three schemes for performance comparison.

\begin{itemize}
	\item Benchmark ignoring CSI errors:
	The AP and WDs optimize the transceiver design via solving problem (P1) by ignoring the channel estimation errors.

	\item Full power transmission:
	Each WD \( k \in \mathcal{K}\) transmits with the full power and aligned phase, i.e., \(b_k = \sqrt{P} \frac{\boldsymbol{\widehat{h}}_k^H \boldsymbol{w}}{|\boldsymbol{w}^H \boldsymbol{\widehat{h}}_k|}\).
	
	\item Channel inversion power control:
	Each WD \( k \in \mathcal{K}\) sets its transmit coefficient based on channel inversion, i.e., \(b_k = \sqrt{P} \frac{\min_{i \in \mathcal{K}}\|\boldsymbol{\widehat{h}}_i\|}{\|\boldsymbol{\widehat{h}}_k\|} \frac{\boldsymbol{\widehat{h}}_k^H \boldsymbol{w}}{|\boldsymbol{w}^H \boldsymbol{\widehat{h}}_k|}\), where the WD with the poorest channel uses up its transmit power.  		
\end{itemize}
Notice that in the last two schemes, the receive beamforming vector \(\boldsymbol{w}\) is designed similarly as in Sections III and IV for the cases with SISO and SIMO, respectively.  In the simulation, we set the channel vectors \(\{\boldsymbol{h}_k\}\) as independent CSCG random vectors with zero mean and covariance \(\sigma_{h,k}^2 \boldsymbol{I}\), where \(\sigma_{h,k}^2\) is generated randomly to capture the differences of large-scale fading at different WDs. We also set \(\sigma_{e,k}^2 = \sigma_{e}^2\) and \(P_k = P\), \(\forall k \in \mathcal{K}\).

\begin{figure}[tb]
	\centering
	{\includegraphics[width=0.32\textwidth]{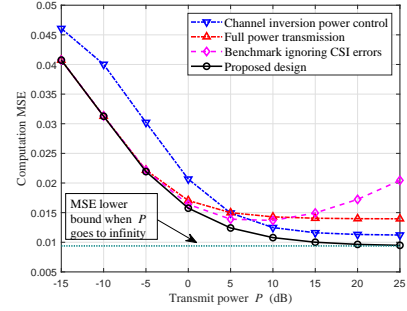}}
	\caption{The computation MSE versus transmit power \(P\) when \(N_r = 1\), \(K = 20\), and \(\sigma_{e}^2 = 0.1\).}
\end{figure}

\begin{figure}[tb]
	\centering
	{\includegraphics[width=0.32\textwidth]{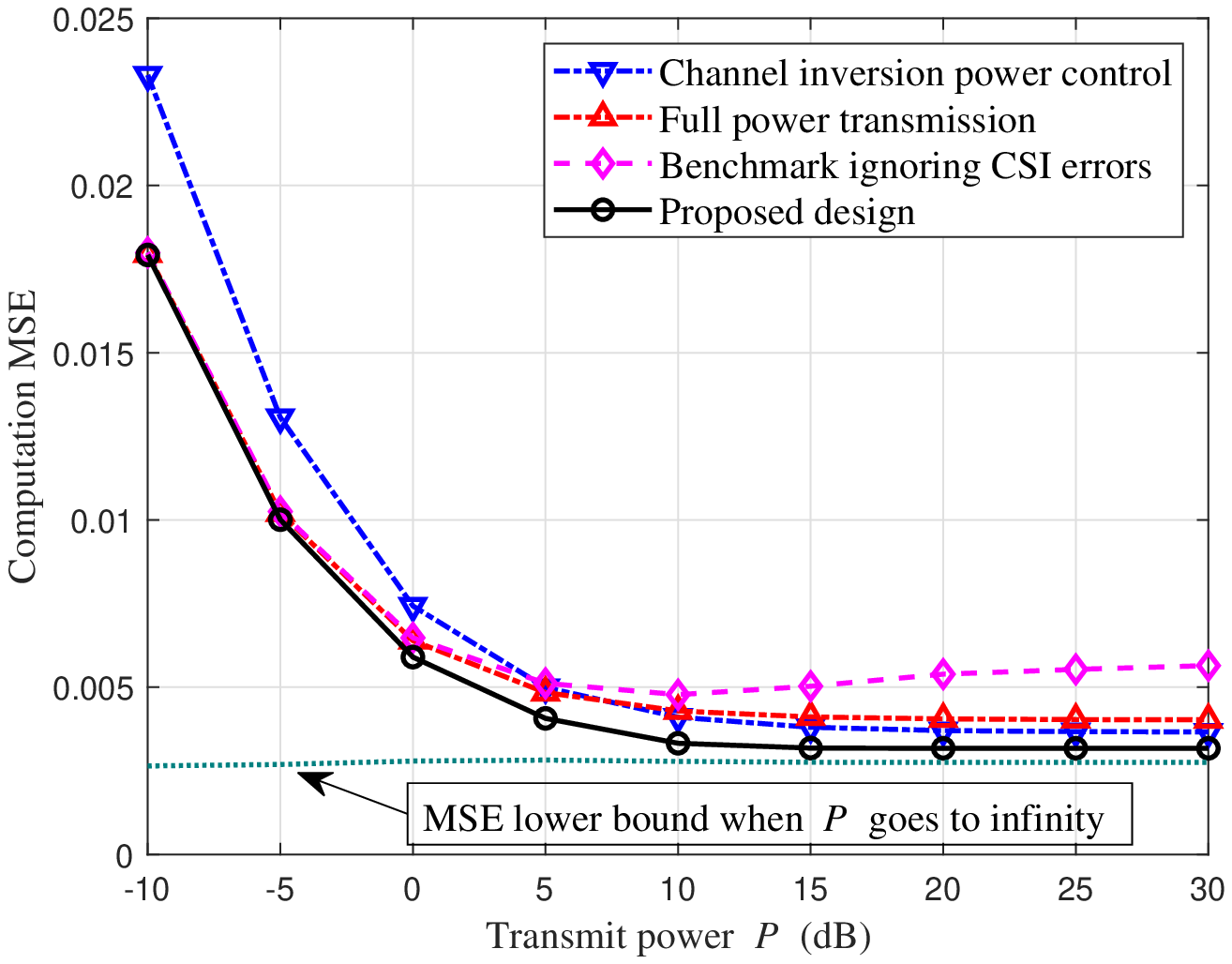}}
	\caption{The computation MSE versus transmit power \(P\) when \(N_r = 10\), \(K = 20\), and \(\sigma_{e}^2 = 0.1\).}
\end{figure}

\begin{figure*}[tb]
	\centering
	\begin{minipage}{0.32\textwidth}
		{\includegraphics[width=1\textwidth]{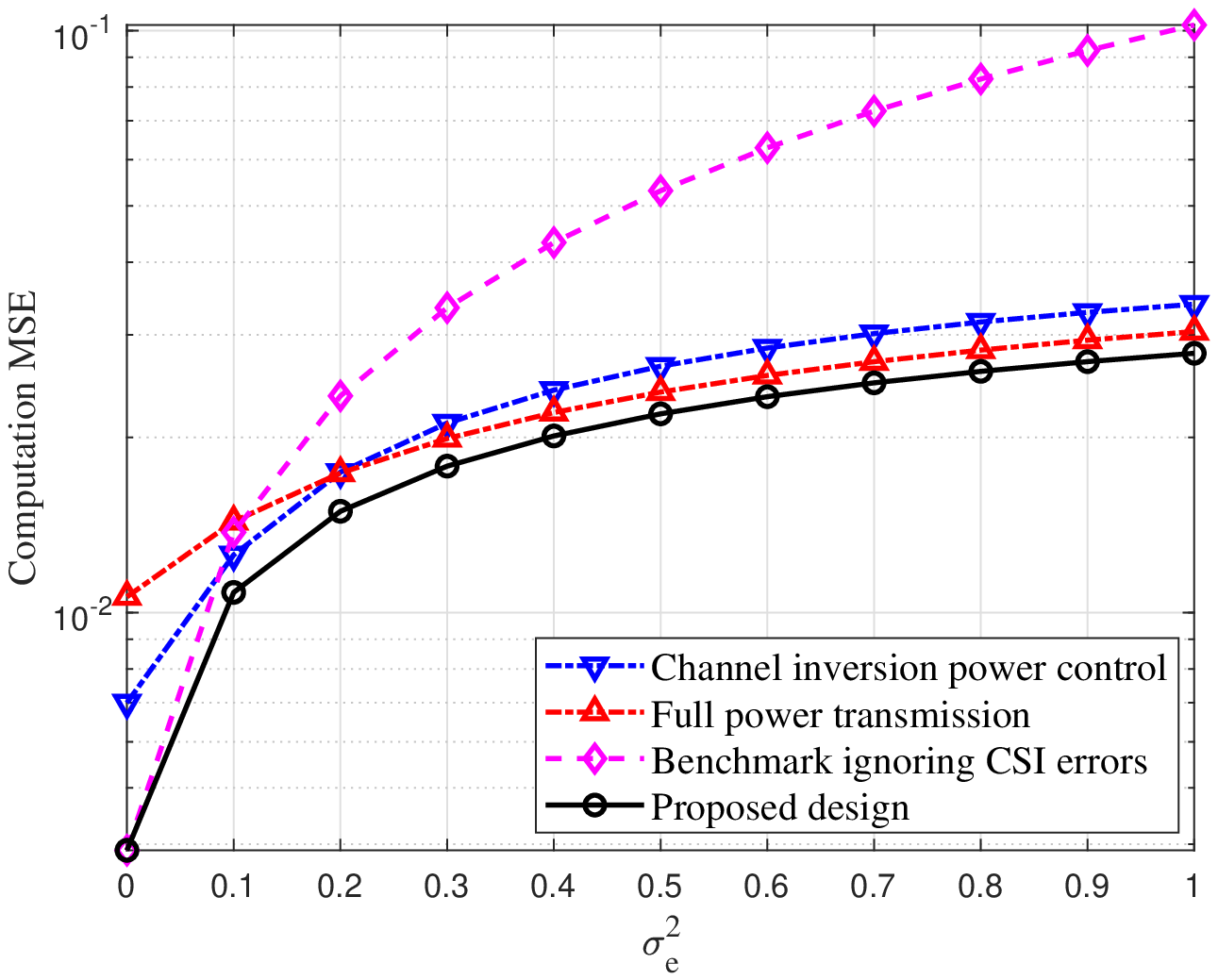}}
		\caption{The computation MSE versus \(\sigma_{e}^2\) when \(N_r = 1\), \(P = 10\) dB, and \(K = 20\).}
	\end{minipage}
	\begin{minipage}{0.32\textwidth}
		{\includegraphics[width=1\textwidth]{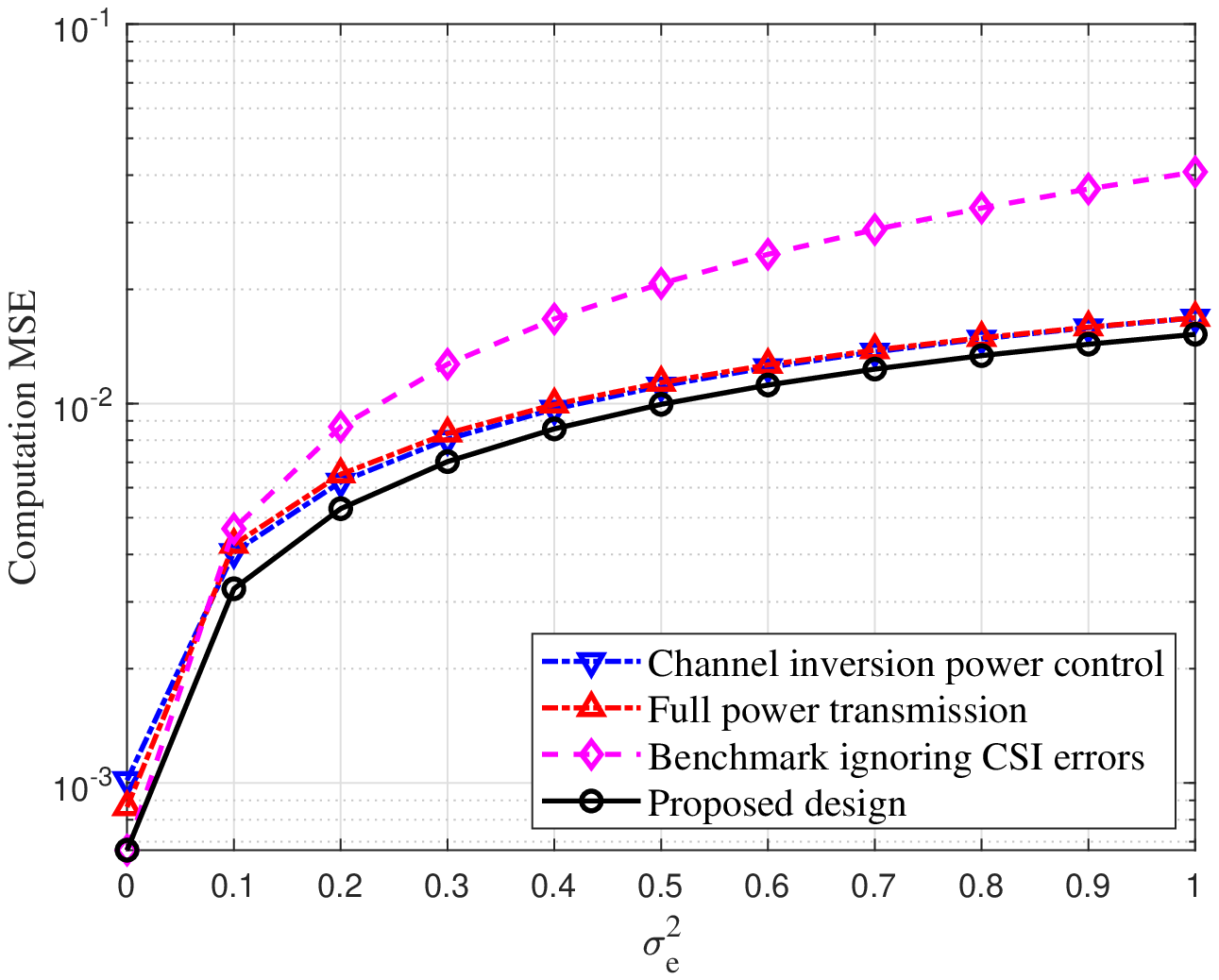}}
		\caption{The computation MSE versus \(\sigma_{e}^2\) when \(N_r = 10\), \(P = 10\) dB, and \(K = 20\).}
	\end{minipage}
	\begin{minipage}{0.32\textwidth}
		\includegraphics[width=1\textwidth]{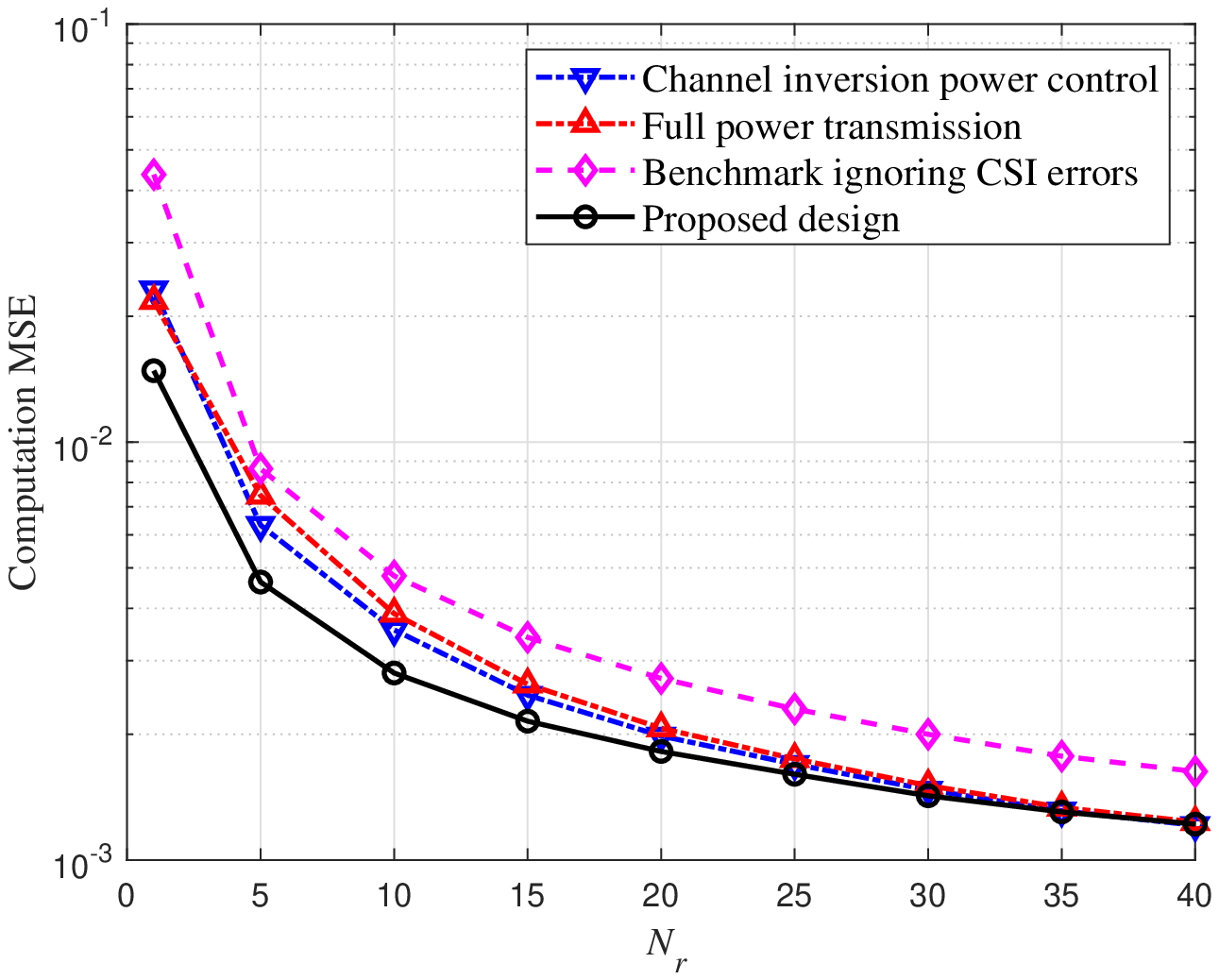}
		\caption{The computation MSE versus \(N_r\) when \(P = 10\) dB, \(K = 20\), and \(\sigma_{e}^2 = 0.1\).}
	\end{minipage}
\end{figure*}

Figs. 1 and 2 show the computation MSE versus the transmit power \(P\) at each WD for the cases with SISO and SIMO (\(N_r = 10\)), respectively, where \(K = 20\) and \(\sigma_{e}^2 = 0.1\). It is observed that the proposed design outperforms the other benchmarks in the whole transmit power regime. In the low power regime (e.g., \(P \le 0\) dB), the full power transmission is observed to perform close to the proposed design, as it can efficiently suppress the noise-induced error that is dominant in MSE in this case. In the high power regime (e.g., \(P \ge 15\) dB), the channel inversion power control is observed to perform close to the proposed design, due to the efficient signal magnitude alignment. When \(P\) becomes large, the computation MSE achieved by the proposed design is observed to approach the lower bound, as indicated in Propositions 1 and 2. By contrast, the benchmark ignoring CSI errors is observed to perform the worst and even leads to an increased MSE when \(P\) becomes large. This is because the CSI errors are amplified by the high transmit power, hence degrading the MSE performance.

Figs. 3 and 4 show the computation MSE versus the variance of CSI error \(\sigma_e^2\) for the cases with SISO and SIMO (\(N_r = 10\)), respectively, where \(K = 20\) and \(P = 10\) dB. It is observed that the proposed design outperforms the other benchmarks in the whole regime of \(\sigma_{e}^2\). The performance gap between the proposed design and the benchmark ignoring CSI errors is observed to become more significant as \(\sigma_e^2\) increases.  


Fig. 5 shows the computation MSE versus the number of receive antennas \(N_r\) at the AP, where \(K = 20\), \(P = 10\) dB, and \(\sigma_e^2 = 0.1\). It is observed that the performances of channel inversion power control and full power transmission gradually approach that of the proposed design as \(N_r\) increases, as the receive beamforming becomes more critical for data aggregation (see Proposition 3), and the gain of transmit power control becomes less. In addition, the benchmark ignoring CSI errors performs worse than the others over the whole regime of \(N_r\).

\section{Conclusion}

This paper considered the joint transceiver design to minimize the computation MSE for an uncoded AirComp system with imperfect CSI. For the SISO case, we derived the optimal threshold-based regularized channel inversion power control solution to the computation MSE minimization problem; while for the SIMO case, we proposed an alternating-optimization-based algorithm to find a high-quality solution. In addition, we derived interesting analytic results on the computation MSE in the regimes with asymptotically high transmit powers or an asymptotically large number of receive antennas. Remarkable MSE performance gains were observed by our proposed designs, in the comparison with benchmark schemes, which showed the importance of jointly optimizing the transmit power control and the receive strategy to combat against the channel estimation errors for reliable AirComp.

\bibliographystyle{IEEEtran}
\bibliography{IEEEabrv,AirComp3}

\end{document}